\documentclass[11pt,a4paper]{article}
\usepackage{amsmath,amssymb}
\usepackage{epsfig,graphicx}
\usepackage{cite}

\begin{document}

\title{
\begin{flushright}
{\normalsize Yaroslavl State University\\
             Preprint YARU-HE-06/02\\
             hep-ph/0606259} \\[20mm]
\end{flushright}
{\bf Neutrino dispersion in external magnetic field\\ and plasma}
}
\author{A.~V.~Kuznetsov$^a$\footnote{{\bf e-mail}: avkuzn@uniyar.ac.ru},
N.~V.~Mikheev$^{a}$\footnote{{\bf e-mail}: mikheev@uniyar.ac.ru}
\\
$^a$ \small{\em Yaroslavl State (P.G.~Demidov) University} \\
\small{\em Sovietskaya 14, 150000 Yaroslavl, Russian Federation}
}
\date{}

\maketitle

\begin{abstract}
Neutrino dispersion properties in an active medium 
consisting of magnetic field and plasma are analysed. 
We consider in detail the contribution of a magnetic field 
into the neutrino self-energy operator $\Sigma (p)$. The results 
for this contribution were contradictory in the previous literature.  
For the conditions of the early
universe where the background medium consists of a charge-symmetric
plasma, the pure magnetic field contribution to the neutrino dispersion
relation is proportional to $(e B)^2$ and thus comparable to the
contribution of the magnetized plasma.
We consider one more hypothetical effect of the active medium 
influence on neutrino properties, the so-called ``neutrino spin light'' 
discussed in the literature. 
We show that this effect has no physical region of 
realization because of the medium influence on photon dispersion. 
\end{abstract}

\vfill

\begin{center}
 {\it Based on the talk presented } \\
 {\it at the XIV International Seminar Quarks'2006, } \\
 {\it St.-Petersburg, Repino, Russia, May 19-25, 2006}
\end{center}


\newpage

\section{Introduction}

The most important event in neutrino physics of the last decades was undeniably 
the solving of the Solar neutrino puzzle, made in the unique experiment 
on the heavy-water detector at the Sudbury Neutrino Observatory. 
This experiment, together with the atmospheric and the reactor neutrino 
experiments, has confirmed the key idea by 
B. Pontecorvo on neutrino oscillations. 
The existence of non-zero neutrino mass and lepton mixing is thereby established. 

In this connection, an enthusiasm has arisen among theorists with respect to 
the searches of other possible detectable effects in neutrino physics.                                       
However, when people are enthusiastic, they could be not enough self-critical.  
Sometimes, theoretical discoveries are followed by theoretical closings. 
In this paper, we have to close two recent discoveries in the field of 
neutrino dispersion in external active medium. 

The first one concerned the external magnetic field influence on 
the neutrino dispersion relation. In the papers by E.~Elizalde 
et al.~\cite{Elizalde:2002,Elizalde:2004}, contrary to 
previous results obtained by several authors, a gigantic 
field contribution into the neutrino energy was found. 
If the result was correct, and if the previous hunters in the field 
really ignored ``the elephant'',
it would lead to important consequences for neutrino physics in media. 

One more promising effect based on using the neutrino dispersion properties 
in external active medium, the so-called ``spin light of neutrino'', 
was proposed in the series of 
papers~\cite{Lobanov:2003,Lobanov:2004,Studenikin:2005,Lobanov:2005,Grigoriev:2005}.  
However, the medium influence on the 
photon dispersion was not considered there. 

\section{Neutrino dispersion in magnetized plasma: a conflict}

The presence of matter or electromagnetic fields modifies the
dispersion relation of neutrinos only slightly because these
particles interact only by the weak force. However,
it was recognized that the feeble matter effect
is enough to affect neutrino flavor oscillations in dramatic ways
because the neutrino mass differences are very
small~\cite{Wolfenstein:1978,Mikheyev:1985}, with practical applications
in physics and astrophysics whenever neutrino oscillations
are important~\cite{Mohapatra:1998rq}.

\def\D{\mathrm{d}} 
\def\E{\mathrm{e}}
\def\I{\mathrm{i}}

The presence of external fields will lead to additional modifications
of the neutrino dispersion relation. There is a natural scale for the
field strength that is required to have a significant impact on
quantum processes, i.e.\ the critical value
\begin{equation}
B_e = m_e^2/e \approx 4.41 \times 10^{13}~\textrm{G}\,.
\end{equation}
Note that we use natural units where $\hbar=c=1$ and the
Lorentz-Heaviside convention where
$\alpha=e^2/4\pi\approx1/137$ so that $e\approx0.30>0$ is the
elementary charge, taken to be positive.

There are reasons to expect that fields of such or even larger
magnitudes can arise in cataclysmic astrophysical events such as
supernova explosions or coalescing neutron stars, situations where a
gigantic neutrino outflow should also be expected.  There are two
classes of stars, i.e.~soft gamma-ray repeaters
(SGR) and anomalous x-ray pulsars
(AXP) that are believed to be remnants
of such cataclysms and to be magnetars, neutron
stars with magnetic fields $10^{14}$--$10^{15}$~G.  The
possible existence of even larger fields of order
$10^{16}$--$10^{17}$~G is subject to
debate, see e.g.~\cite{Ardeljan:2004} and the references cited 
therein.  The early universe between
the QCD phase transition (${}\sim 10^{-5}$~s) and the nucleosynthesis
epoch (${}\sim 10^{-2}$--$10^{+2}$~s) is believed to be yet another
natural environment where strong magnetic fields and large neutrino
densities could exist simultaneously~\cite{Grasso:2001}.

The modification of the neutrino dispersion relation in a magnetized
astrophysical plasma was studied in the previous
literature~\cite{D'Olivo:1989,Semikoz:1994,Elmfors:1996,
Erdas:1998}.  In particular, a charge-symmetric plasma with $m_e \ll T
\ll m_W$ and $B \lesssim T^2$ was considered for the early-universe
epoch between the QCD phase transition and big-bang nucleosynthesis.
Ignoring the neutrino mass, the dispersion relation for the electron
flavor was found to be~\cite{Elmfors:1996,Erdas:1998}
\begin{eqnarray}
\frac{E}{|{\bf p}|}=1&+&
\frac{\sqrt{2}\,G_{\rm F}}{3}
\left[
-\frac{7\pi^2T^4}{15} 
\left(\frac{1}{m_Z^2} + \frac{2}{m_W^2} \right)
+ \frac{T^2eB}{m_W^2}\,\cos\phi \, + \right. 
\nonumber\\
&+& \left. \, \frac{(eB)^2}{2\pi^2m_W^2}\,
\ln\left(\frac{T^2}{m_e^2}\right)\,\sin^2 \phi \right],
\label{eq:E_Raf}
\end{eqnarray}
where $\bf p$ is the neutrino momentum and $\phi$ is the angle between
$\bf B$ and $\bf p$.  The first term proportional to $G_{\rm F}$
in Eq.~(\ref{eq:E_Raf}) is the
dominating pure plasma contribution~\cite{Notzold:1988}, whereas the
second term is caused by the common influence of the plasma and
magnetic field~\cite{Elmfors:1996}.  The third term is of the second order
in $(eB/T^2)\ll 1$ but was included because of the large logarithmic
factor $\ln(T/m_e)\gg 1$ \cite{Erdas:1998}. The dispersion relation
of Eq.~({\ref{eq:E_Raf}}) applies to both $\nu_e$ and $\bar\nu_e$
without sign change in any of the terms.

The $B$-field induced pure vacuum modification of the neutrino
dispersion relation 
was assumed to be negligible in these papers.

However, this contribution was calculated for the same
conditions in Refs.~\cite{Elizalde:2002,Elizalde:2004} 
with an absolutely different result:
\begin{equation}
\frac{\Delta E}{|{\bf p}|}= 
\sqrt{2}\,G_{\rm F}\,
\frac{eB}{8\pi^2}\,\sin^2 \phi \; \E^{-p_\bot^2/(2eB)}\,,
\label{eq:DE_Elizalde}
\end{equation}
where $p_\bot$ is the momentum component perpendicular to the
$B$-field. It is easy to check that this would be the dominant
$B$-field induced contribution by far and thus would lead to important
consequences for neutrino physics in media. 

Because of importance of the question whether the $B$-field contribution 
into the neutrino dispersion relation was dominating or negligible, an 
independent calculation of it was strongly urged.

\section{The neutrino self-energy operator $\Sigma(p)$} 
\label{sec:definition}

A literature search reveals that calculations of the neutrino
dispersion relation in external $B$-fields have a long
history~\cite{McKeon:1981,Borisov:1985,Erdas:1990}. To compare the
different results we introduce the neutrino self-energy operator
$\Sigma (p)$ that is defined in terms of the invariant amplitude for
the neutrino forward scattering on vacuum fluctuations, $\nu \to \nu$, 
by the relation
\begin{equation}
{\cal M}(\nu\to\nu)=-\bar\nu(p)\,\Sigma (p)\,\nu(p)\,,
\label{eq:sigma_def}
\end{equation}
where $p$ is the neutrino four-momentum. Note that we use the
signature $(+,-,-,-)$ for the four-metric. 

Perturbatively, the matrix element of Eq.~(\ref{eq:sigma_def})
corresponds to the one-loop $\nu \to \ell + W \to \nu$ transition 
with the exact propagators taken for $\ell$ and $W$ in the external $B$ field.
The calculation techniques for loop processes in external electromagnetic fields
based on exact propagators started from the classical paper by 
J. Schwinger~\cite{Schwinger:1951} and was developed by A.~Nikishov, V. Ritus, A. Shabad, 
V. Skobelev et al. For a recent review see e.g.~\cite{Kuznetsov:2003}. 

It is convenient to express the structure of the $\Sigma(p)$ operator 
in an external magnetic field in terms of the coefficients 
${\cal A}_L$, ${\cal B}_L$, ${\cal C}_L$, ${\cal A}_R$, etc.

\begin{eqnarray}
\Sigma(p)&=&
\left[{\cal A}_L\,(p\gamma) + {\cal B}_L \,e^2 \left(p \tilde F \tilde F \gamma \right) + 
{\cal C}_L \,e \left(p \tilde F \gamma \right) \right]\, L 
\nonumber\\[2mm]
&+& \left[{\cal A}_R\,(p\gamma) + {\cal B}_R\,e^2 \left(p \tilde F \tilde F \gamma \right) + 
{\cal C}_R \, e \left(p \tilde F \gamma \right) \right]\, R 
\nonumber\\[2mm]
&+& 
m_\nu \, \left[{\cal K}_1 + \I \, {\cal K}_2 \, e \left(\gamma F \gamma \right) \right]
\,, 
\label{eq:sigma_cros}
\end{eqnarray}
where $F$ is the external field tensor, and $\tilde F$ is its dual, 
$L=\frac{1}{2}(1-\gamma_5)$, $R=\frac{1}{2}(1+\gamma_5)$. 

Here, the coefficients ${\cal A}_L$, ${\cal A}_R$ and ${\cal K}_1$, 
being ultraviolet divergent, do not have independent meanings, 
because they do not give contributions into the real neutrino energy in external field 
at the one-loop level. 
They are absorbed by the neutrino wave-function and mass renormalization.
The coefficients ${\cal B}_R$, ${\cal C}_R$ are suppressed by the factor $(m_{\nu}/m_W)^2$. 
The coefficient ${\cal K}_2$ is suppressed by the factor $(m_{\ell}/m_W)^2$. 
Thus, the coefficients ${\cal B}_L$, ${\cal C}_L$ are of the most interest. 
The collection of the results for the ${\cal B}_L$ and ${\cal C}_L$ coefficients 
of the $\Sigma(p)$ operator~(\ref{eq:sigma_cros}) is presented in our 
paper~\cite{Kuznetsov:2006}. 
Our results for relatively weak field $e B \ll m_\ell^2 \ll m_W^2$ are
\begin{equation}
{\cal B}_L = - \frac{G_{\rm F}}{3 \sqrt{2}\,\pi^2 \, m_W^2}
\left(\ln\frac{m_W^2}{m_\ell^2}+\frac{3}{4}\right), \qquad
{\cal C}_L = \frac{3 G_{\rm F}}{4 \sqrt{2}\,\pi^2}
\,.
\label{eq:coef_weak}
\end{equation}
For moderate field $m_\ell^2 \ll e B \ll m_W^2$ we have obtained
\begin{equation}
{\cal B}_L = - \frac{G_{\rm F}}{3 \sqrt{2}\,\pi^2 \, m_W^2}
\left(\ln\frac{m_W^2}{eB}+2.54\right), 
\label{eq:coef_moder}
\end{equation}
with the same coefficient ${\cal C}_L$. 

\section{Neutrino energy in a magnetic field}
\label{sec:energy}

Solving the equation for the neutrino dispersion in a magnetic field
($m_\nu \equiv 0$)

\begin{equation}
\textrm{det} \, \left|(p \gamma) - {\cal B}_L \, e^2 (p \tilde F \tilde F \gamma) \,L
- {\cal C}_L \, e (p \tilde F \gamma) \,L \right| = 0 
\,,
\label{eq:det}
\end{equation}
where the leading terms with ${\cal B}_L$, ${\cal C}_L$
are only included, one obtains for the neutrino energy in the field:

\begin{equation}
\frac{E}{|{\bf p}|} = 1 +
\left({\cal B}_L + \frac{{\cal C}_L^2}{2}\right) (e B)^2 \, \sin^2 \phi \,.
\label{eq:E/p_CB}
\end{equation}

It can be seen that the ${\cal B}_L$ coefficient gives the main contribution into 
the neutrino energy, because the value ${\cal C}_L^2/{\cal B}_L \sim G_{\rm F} m_W^2$ 
appears to be of the 
order of the fine-structure constant $\alpha \simeq 1/137$, thus leading us 
beyond the frame of the one-loop approximation.

Our results strongly disagree with those by E.~Elizalde 
et al.~\cite{Elizalde:2002,Elizalde:2004}. We think that the
disagreement arises because these authors use only one lowest Landau
level in the charged-lepton propagator in the case of moderate field
strengths which they call ``strong fields.''  However, the
contributions of the next Landau levels appear to be of the same order as
the ground-level contribution~\cite{Kuznetsov:2006} because in the integration over the
virtual lepton four-momentum in the loop the region $q^2 \sim
m_W^2 \gg e B$ appears to be essential.  

We confirm the assumption~\cite{D'Olivo:1989,Elmfors:1996},
that the pure magnetic field contribution into the neutrino energy does not exceed the 
plasma contribution.

For relatively weak field $e B \ll m_e^2$ we find the pure-field correction to 
the electron neutrino energy in a magnetic field and plasma, rewriting the last 
term in Eq.~(\ref{eq:E_Raf}) to the form:
\begin{eqnarray}
+ \frac{(e B)^2}{2\pi^2m_W^2}
\,\sin^2 \phi 
\left( \ln \frac{T^2}{m_e^2} - \ln \frac{m_W^2}{m_e^2} - 
\frac{3}{4}\right) \,.
\label{eq:E/p_final}
\end{eqnarray}

It is seen that 
the pure magnetic field contribution to the neutrino dispersion 
is proportional to $(e B)^2$ and thus comparable to the 
contribution of the magnetized plasma. It is interesting to note 
that the contributions of plasma and of pure magnetic field 
into Eq.~(\ref{eq:E/p_final}), containing the 
electron mass singularities $\sim \ln m_e$, exactly cancel each other. 

\section{No ``neutrino spin light'' because of photon dispersion in medium}
\label{sec:spinlight}

In an astrophysical environment, the main medium influence 
on neutrino properties is defined by the additional Wolfenstein 
energy $W$ acquired by a left-handed neutrino~\cite{Wolfenstein:1978}. 

The general expression for this additional energy of 
a left-handed neutrino with the flavor $i = e, \mu, \tau$ 
is~\cite{Notzold:1988,Pal:1989,Nieves:1989}
\begin{eqnarray}
&&W_i = \sqrt{2} \, G_{\rm F} \left[
\left(\delta_{ie} - \frac{1}{2} + 2 \, \sin^2 \theta_{\rm W}\right) 
\left(N_e - \bar N_e \right) \right.
\label{eq:EnuLgen}\\
&&+ \left. \left(\frac{1}{2} - 2 \, \sin^2 \theta_{\rm W}\right) 
\left(N_p - \bar N_p \right) 
- \frac{1}{2} \left(N_n - \bar N_n \right) 
+ \frac{1}{2} \, \sum\limits_{\ell = e, \mu, \tau} 
\left(N_{\nu_\ell} - \bar N_{\nu_\ell} \right) 
\right] , 
\nonumber
\end{eqnarray}
where the functions 
$N_e, N_p, N_n, N_{\nu_\ell}$ are the number densities of background electrons, 
protons, neutrons, and neutrinos, and $\bar N_e, \bar N_p, \bar N_n, 
\bar N_{\nu_\ell}$ are the densities of the antiparticles. 
To find the additional energy for antineutrinos, one should change the 
total sign in the right-hand side of Eq.~(\ref{eq:EnuLgen}). 

For a typical astrophysical medium, except for the early Universe and 
a supernova core, 
one has $\bar N_e \simeq \bar N_p \simeq \bar N_n \simeq N_{\nu_\ell}
\simeq \bar N_{\nu_\ell} \simeq 0$,
and $N_p \simeq N_e = Y_e\, N_B, \, N_n \simeq  (1-Y_e)\, N_B$, where $N_B$ 
is the barion density. One obtains
\begin{eqnarray}
W_e = \frac{G_{\rm F} \, N_B}{\sqrt{2}} 
\left(3\, Y_e-1 \right) , \qquad
W_{\mu,\tau} = - \frac{G_{\rm F} \, N_B}{\sqrt{2}} 
\left(1-Y_e \right) .
\label{eq:EnuL}
\end{eqnarray}
As $Y_e < 1$, the additional energy acquired by muon and tau left-handed 
neutrinos 
is always negative. At the same time, the additional energy of electron 
left-handed neutrinos becomes positive at $Y_e > 1/3$. 
And vice versa, the additional energy for electron antineutrinos is positive 
at $Y_e < 1/3$, while it is always positive for the muon and tauon 
antineutrinos. 
On the other hand, right-handed neutrinos and their antiparticles, left-handed 
antineutrinos, being sterile with respect to weak interactions, 
do not acquire an additional energy.

The additional energy $W$ from Eq.~(\ref{eq:EnuL}) 
gives an effective mass squared $m_L^2$ to the left-handed neutrino, 
\begin{equation}
m_L^2 = {\cal P}^2 = (E + W)^2 - {\bf p}^2 \,, 
\label{eq:mL}
\end{equation}
where ${\cal P}$ is the neutrino four-momentum in medium, while 
$(E,\, {\bf p})$ would form the neutrino four-momentum in vacuum, 
$E = \sqrt{{\bf p}^2 + m_\nu^2}$. 

Given a $\nu \nu \gamma$ interaction, 
the additional energy of left-handed neutrinos in medium opens new 
kinematical possibilities for the radiative neutrino transition 
$\nu \to \nu + \gamma$.
It should be self-evident, that the influence of the substance on the photon 
dispersion must be taken into account, 
$\omega = |{\bf k}|/n$, where $n \ne 1$ is the refractive index. 

First, a possibility exists that the medium provides the condition $n > 1$ 
(the effective photon mass squared is negative, $m_\gamma^2 \equiv 
q^2 < 0$) which corresponds to the well-known 
effect~\cite{Grimus:1993,D'Olivo:1996,Ioannisian:1997} of 
``neutrino Cherenkov radiation''.  
In this situation, the neutrino dispersion change under the medium influence 
is being usually neglected, because the neutrino dispersion is defined by the 
weak interaction while the photon dispersion is defined by the electromagnetic 
interaction. 

Pure theoretically, one more possibility could be considered when the photon 
dispersion was absent, and the process of the radiative neutrino transition
$\nu \to \nu \gamma$ would be caused by the neutrino dispersion only. 
As the left-handed neutrino dispersion is only changed, transitions become 
possible caused by the $\nu \nu \gamma$ interaction with the neutrino 
chirality change, e.g. due to the neutrino magnetic dipole moment. 

Just this situation called the ``spin light of neutrino'' ($SL \nu$), 
was first proposed and investigated in detail in an extended series of 
papers~\cite{Lobanov:2003,Lobanov:2004,Studenikin:2005,Lobanov:2005,Grigoriev:2005}.  
However, in the analysis of this effect the authors overlooked such an 
important phenomenon as plasma influence on the photon dispersion. 
As will be shown below, this phenomenon closes the $SL \nu$ effect for all 
real astrophysical situations. 

We have reanalysed the process 
$\nu_L \to \nu_R \gamma$ taking into account both the neutrino 
dispersion and the photon dispersion in medium. 
Having in mind possible astrophysical applications, it is worthwhile to consider 
the astrophysical plasma as a medium, which transforms the photon into 
the plasmon, see e.g. Ref.~\cite{Braaten:1993} and the papers cited 
therein. 

To perform a kinematical analysis, it is necessary to evaluate the scales 
of the values of the left-handed neutrino additional energy $W$ and of the 
photon (plasmon) effective mass squared $m_\gamma^2$.
One readily obtains from Eq.~(\ref{eq:EnuL}) for the electron neutrino: 
\begin{equation}
W \simeq 6 \; {\rm eV}
\left(\frac{N_B}{10^{38} \, {\rm cm}^{-3}}\right) \left(3\, Y_e - 1 \right) ,
\label{eq:W}
\end{equation}
where the scale of the barion number density is taken, which is typical 
e.g. for the interior of a neutron star. 

On the other hand, a plasmon acquires in medium 
an effective mass $m_\gamma$ which is approximately constant at high energies. 
For the transversal plasmon, the value $m_\gamma^2$ is always positive, and 
is defined by the so-called plasmon frequency. 
In the non-relativistic classical plasma (i.e. for the solar interior) one has:
\begin{equation}
m_\gamma \equiv \omega_{\rm pl} = \sqrt{\frac{4 \pi \, \alpha \,N_e}{m_e}} \simeq 
4 \times 10^{2} \,{\rm eV}
\left(\frac{N_e}{10^{26} {\rm cm}^{-3}}\right)^{1/2}.
\label{eq:omega_pl_nr}
\end{equation}
For the ultra-relativistic dense matter one has:
\begin{equation}
m_\gamma =  \sqrt{\frac{3}{2}} \; \omega_{\rm pl} 
= \left(\frac{2 \, \alpha}{\pi} \right)^{1/2} 
\left(3\, \pi^2 \, N_e \right)^{1/3} \simeq 
10^{7} \,{\rm eV}
\left(\frac{N_e}{10^{37} \, {\rm cm}^{-3}}\right)^{1/3}.
\label{eq:omega_pl_r}
\end{equation}
In the case of hot plasma, when its temperature is the largest 
physical parameter, the plasmon mass is:
\begin{equation}
m_\gamma =  \sqrt{\frac{2 \, \pi \, \alpha}{3}} \; T
\simeq 
1.2 \times 10^{7} \,{\rm eV}
\left(\frac{T}{100 \, {\rm MeV}}\right).
\label{eq:omega_pl_h}
\end{equation}

One more physical parameter, a great attention was payed to in the $SL \nu$ 
analysis~\cite{Lobanov:2003,Lobanov:2004,Studenikin:2005,Lobanov:2005,Grigoriev:2005},
was the neutrino vacuum mass $m_\nu$. 
As the scale of neutrino vacuum mass could not exceed essentially 
a few electron-volts, which 
is much less than typical plasmon mass scales for real astrophysical situations, 
see Eqs.~(\ref{eq:omega_pl_nr})-(\ref{eq:omega_pl_h}), it is reasonable 
to neglect $m_\nu$ in our analysis.

Thus, in accordance with~(\ref{eq:mL}), a simple condition for the 
kinematic opening of the process $\nu_L \to \nu_R \gamma$ is:
\begin{equation}
m_L^2 \simeq 2 \, E \, W > m_\gamma^2. 
\label{eq:cond}
\end{equation}
This means that the process becomes kinematically opened when 
the neutrino energy exceeds the threshold value, 
\begin{equation}
E > E_0 = m_\gamma^2/(2 \, W). 
\label{eq:threshold}
\end{equation}

Let us evaluate these threshold neutrino energies for different astrophysical 
situations. 
For the solar interior $N_B \simeq 0.9 \times 10^{26} \, {\rm cm}^{-3}$, 
$Y_e \simeq 0.6$, and the threshold neutrino energy is:
$E_0 \simeq 10^{10} \,{\rm MeV}$, 
to be compared with the upper bound $\sim$ 20 MeV for the solar neutrino energies. 

For the interior of a neutron star, where $Y_e \ll 1$, 
the Wolfenstein energy for neutrinos~(\ref{eq:EnuL}) 
is negative, and the process $\nu_L \to \nu_R \gamma$ is closed. 
On the other hand, there exists a possibility for opening the antineutrino 
decay. Taking for the estimation $Y_e \simeq 0.1$, one obtains from~(\ref{eq:W}) 
and~(\ref{eq:omega_pl_r}) the threshold value:
$E_0 \simeq 10^{7} \,{\rm MeV}$, 
to be compared with the typical energy $\sim$ MeV of neutrinos emitted 
via the URCA processes. 

For the conditions of a supernova core, the additional energy of left-handed 
electron neutrinos can be obtained from Eq.~(\ref{eq:EnuLgen}) as follows:
\begin{equation}
W_e = \frac{G_{\rm F} \, N_B}{\sqrt{2}} 
\left(3\, Y_e + Y_{\nu_e} - 1 \right) ,
\label{eq:EnuLeSN}
\end{equation}
where $Y_{\nu_e}$ describes the fraction of trapped electron neutrinos 
in the core, $N_{\nu_e} = Y_{\nu_e}\, N_B$. 
Taking typical parameters of a supernova core, we obtain:
$E_0 \simeq 10^{7} \,{\rm MeV}$, 
to be compared with the averaged energy $\sim 10^{2}$ MeV of trapped neutrinos. 

In the early Universe, when plasma was almost charge symmetric, 
the Wolfenstein formula~(\ref{eq:EnuLgen}) giving zero should be changed 
to a more accurate expression for the additional energy which is identical 
for both neutrinos and antineutrinos~\cite{Notzold:1988,Elmfors:1996} 
\begin{equation}
W_i = - \frac{7 \, \sqrt{2} \, \pi^2 \, G_{\rm F} \, T^4}{45} 
\left( \frac{1}{m_Z^2} + \frac{2 \, \delta_{ie}}{m_W^2} \right) E \, .
\label{eq:W_early}
\end{equation}
The minus sign unambiguously shows that in the early Universe, in contrast 
to the neutron star interior, the decay process is forbidden 
both for neutrinos and antineutrinos. 

Thus, the above analysis shows that the nice effect of the 
``neutrino spin light'', unfortunately, has no place in real 
astrophysical situations because of the photon dispersion. 
The sole possibility for the discussed process $\nu_L \to \nu_R \gamma$ 
to have any significance could be connected only with the situation 
when an ultra-high energy neutrino threads a star. 
Obviously it could have only a methodical meaning. 
The result of a correct calculation of the process width for these purposes 
will be published elsewhere. 

\section{Conclusions}
\label{sec:conclusions}

\begin{itemize}
\item
We have calculated the neutrino self-energy operator $\Sigma (p)$ in the
presence of a magnetic field $B$. 
Our results strongly disagree with those by E.~Elizalde 
et al.~\cite{Elizalde:2002,Elizalde:2004}.
We confirm the assumption by J.~C.~D'Olivo e.a.~\cite{D'Olivo:1989} and by 
P.~Elmfors e.a.~\cite{Elmfors:1996}, 
that the pure magnetic field contribution into the neutrino energy does not exceed the 
plasma contribution.
\item
We have shown that the effect of ``neutrino spin 
light''~\cite{Lobanov:2003,Lobanov:2004,Studenikin:2005,Lobanov:2005,Grigoriev:2005}  
has no physical region of realization because of the photon dispersion 
in medium. 
\end{itemize}

\section*{Acknowledgements}

The authors are grateful to G.~G.~Raffelt and L.~A.~Vassilevskaya for collaboration 
and to V.~A.~Rubakov for useful discussion. 
We express our deep gratitude to the organizers of the 
Seminar ``Quarks-2006'' for warm hospitality.

The work was supported in part 
by the Russian Foundation for Basic Research under the Grant No. 04-02-16253, 
and by the Council on Grants by the President of Russian Federation 
for the Support of Young Russian Scientists and Leading Scientific Schools of 
Russian Federation under the Grant No. NSh-6376.2006.2.



\end{document}